\documentclass{elsart}

\usepackage[english]{babel}
\usepackage{amsmath}
\usepackage{amssymb}
\usepackage{graphicx}

\newcommand{\eg} {{\it e.g.}}

\newcommand{\pnu}[1] {\overset{\smash{\scriptscriptstyle (-)}}{\nu}_{\hskip-3pt #1}}

\begin{document}

\raisebox{8mm}[0pt][0pt]{\hspace{12cm}\vbox{hep-ph/0107150\\IFIC/01-35}}

\begin{frontmatter}

  
  \title{Cornering (3+1) sterile neutrino schemes}


\author{M.\ Maltoni},
\ead{maltoni@ific.uv.es}
\author{T.\ Schwetz} and
\ead{schwetz@ific.uv.es}
\author{J.~W.~F.\ Valle}
\ead{valle@ific.uv.es}

\address{Instituto de F\'{\i}sica Corpuscular -- C.S.I.C., 
  Universitat de Val{\`e}ncia \\
  Edificio Institutos de Paterna, Apt.\ 22085, E--46071 Val{\`e}ncia, Spain}

\begin{abstract}
  Using the most recent atmospheric neutrino data, as well as
  short-baseline, long-baseline and tritium $\beta$-decay data we show that the
  joint interpretation of the LSND, solar and atmospheric neutrino anomalies 
  in (3+1) sterile neutrino schemes is severely disfavored, in contrast
  to the theoretically favored (2+2) schemes.
\end{abstract}

\end{frontmatter}

\section{Introduction}

Reconciling the existing data on solar~\cite{sun-exp} and
atmospheric~\cite{SK-atm-98,atm-exp} neutrinos with a possible hint at
the LSND experiment~\cite{LSND,LSND2000} (indicating the existence of
$\pnu{\mu} \to \pnu{e}$ transitions) is a challenge to the simplest
standard model picture. In the absence of exotic mechanisms and/or new
neutrino interactions, such as neutrino transition magnetic
moment~\cite{Miranda:2001}, one requires the existence of neutrino
oscillations involving three different scales. As a result a joint
explanation for {\it all} the data (including the LSND anomaly)
requires a fourth light neutrino which, in view of the LEP results,
must be sterile~\cite{ptv-pv,cm93} \footnote{This was originally
  postulated to provide some hot dark matter suggested by early COBE
  results.}.

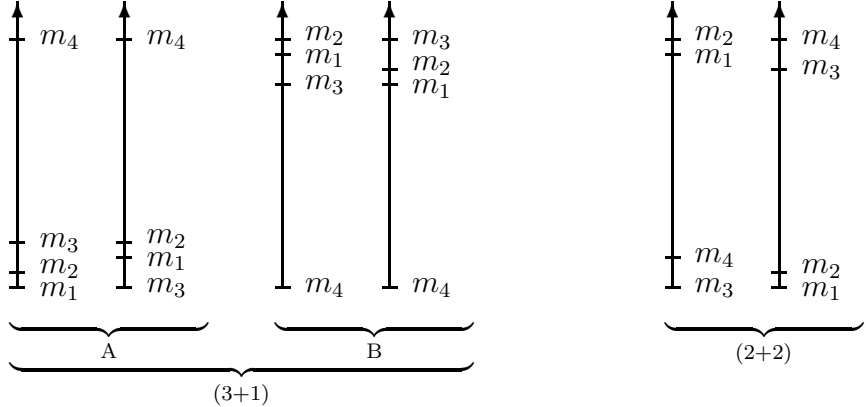
\begin{figure}[t] \centering
    \setlength{\unitlength}{1.0cm}
    $\underbrace{
    \underbrace{
    \begin{tabular}{@{\extracolsep{\fill}}cc}
    \begin{picture}(1,4) 
    \thicklines
    \put(0.1,0.2){\vector(0,1){3.8}}
    \put(0.0,0.2){\line(1,0){0.2}}
    \put(0.4,0.15){\makebox(0,0)[l]{$m_1$}}
    \put(0.0,0.4){\line(1,0){0.2}}
    \put(0.4,0.45){\makebox(0,0)[l]{$m_2$}}
    \put(0.0,0.8){\line(1,0){0.2}}
    \put(0.4,0.8){\makebox(0,0)[l]{$m_3$}}
    \put(0.0,3.5){\line(1,0){0.2}}
    \put(0.4,3.5){\makebox(0,0)[l]{$m_4$}}
    \end{picture}
    &
    \begin{picture}(1,4) 
    \thicklines
    \put(0.1,0.2){\vector(0,1){3.8}}
    \put(0.0,0.2){\line(1,0){0.2}}
    \put(0.4,0.2){\makebox(0,0)[l]{$m_3$}}
    \put(0.0,0.6){\line(1,0){0.2}}
    \put(0.4,0.55){\makebox(0,0)[l]{$m_1$}}
    \put(0.0,0.8){\line(1,0){0.2}}
    \put(0.4,0.85){\makebox(0,0)[l]{$m_2$}}
    \put(0.0,3.5){\line(1,0){0.2}}
    \put(0.4,3.5){\makebox(0,0)[l]{$m_4$}}
    \end{picture}
    \end{tabular}}_{\mathrm{A}}
    \qquad
    \underbrace{
    \begin{tabular}{@{\extracolsep{\fill}}cc}
    \begin{picture}(1,4) 
    \thicklines
    \put(0.1,0.2){\vector(0,1){3.8}}
    \put(0.0,0.2){\line(1,0){0.2}}
    \put(0.4,0.2){\makebox(0,0)[l]{$m_4$}}
    \put(0.0,2.9){\line(1,0){0.2}}
    \put(0.4,2.9){\makebox(0,0)[l]{$m_3$}}
    \put(0.0,3.3){\line(1,0){0.2}}
    \put(0.4,3.25){\makebox(0,0)[l]{$m_1$}}
    \put(0.0,3.5){\line(1,0){0.2}}
    \put(0.4,3.55){\makebox(0,0)[l]{$m_2$}}
    \end{picture}
    &
    \begin{picture}(1,4) 
    \thicklines
    \put(0.1,0.2){\vector(0,1){3.8}}
    \put(0.0,0.2){\line(1,0){0.2}}
    \put(0.4,0.2){\makebox(0,0)[l]{$m_4$}}
    \put(0.0,2.9){\line(1,0){0.2}}
    \put(0.4,2.85){\makebox(0,0)[l]{$m_1$}}
    \put(0.0,3.1){\line(1,0){0.2}}
    \put(0.4,3.15){\makebox(0,0)[l]{$m_2$}}
    \put(0.0,3.5){\line(1,0){0.2}}
    \put(0.4,3.5){\makebox(0,0)[l]{$m_3$}}
    \end{picture}
    \end{tabular}}_{\mathrm{B}}}_{(3+1)}
    \qquad\qquad\qquad
    \underbrace{
    \begin{tabular}{@{\extracolsep{\fill}}cc}
    \begin{picture}(1,4) 
    \thicklines
    \put(0.1,0.2){\vector(0,1){3.8}}
    \put(0.0,0.2){\line(1,0){0.2}}
    \put(0.4,0.2){\makebox(0,0)[l]{$m_3$}}
    \put(0.0,0.6){\line(1,0){0.2}}
    \put(0.4,0.6){\makebox(0,0)[l]{$m_4$}}
    \put(0.0,3.3){\line(1,0){0.2}}
    \put(0.4,3.25){\makebox(0,0)[l]{$m_1$}}
    \put(0.0,3.5){\line(1,0){0.2}}
    \put(0.4,3.55){\makebox(0,0)[l]{$m_2$}}
    \end{picture}
    &
    \begin{picture}(1,4) 
    \thicklines
    \put(0.1,0.2){\vector(0,1){3.8}}
    \put(0.0,0.2){\line(1,0){0.2}}
    \put(0.4,0.15){\makebox(0,0)[l]{$m_1$}}
    \put(0.0,0.4){\line(1,0){0.2}}
    \put(0.4,0.45){\makebox(0,0)[l]{$m_2$}}
    \put(0.0,3.1){\line(1,0){0.2}}
    \put(0.4,3.1){\makebox(0,0)[l]{$m_3$}}
    \put(0.0,3.5){\line(1,0){0.2}}
    \put(0.4,3.5){\makebox(0,0)[l]{$m_4$}}
    \end{picture}
    \end{tabular}}_{(2+2)}$
  \caption{\label{fig:4spectra}%
    The six types of 4--neutrino mass spectra. The different distances between
    the masses on the vertical axes symbolize the different scales of
    mass-squared differences required to explain solar, atmospheric and LSND
    data with neutrino oscillations.}
\end{figure}

There have been several theoretical models and phenomenological
studies of 4--neutrino models~\cite{4-early,Liu:1998,Hirsch:2000a}.
Two very different classes of 4--neutrino mass spectra can be
identified: the first class contains four types and consists of
spectra where three neutrino masses are clustered together, whereas
the fourth mass is separated from the cluster by the mass gap needed
to reproduce the LSND result; the second class has two types where one
pair of nearly degenerate masses is separated by the LSND gap from the
two lightest neutrinos. These two classes will be referred to as (3+1)
and (2+2) neutrino mass spectra, respectively~\cite{barger00}. All
possible 4--neutrino mass spectra are shown in
Fig.~\ref{fig:4spectra}.

Theoretically the existence of a light sterile neutrino sets a
challenge. One possibility is to postulate a protecting
symmetry~\cite{ptv-pv} such as lepton number. Alternatively, in models
based on extra dimensions, one may appeal to a volume suppression
factor~\cite{Ioannisian:2001mu} in order to account for the light
sterile neutrino.  The theoretical origin of the splittings depends on
the model. In Ref.~\cite{Hirsch:2000a} R-parity violating interactions
are used, while in the original proposals the splittings were due to
calculable two-loop effects.
Such theories lead to a (2+2) symmetric scheme where two of the
neutrinos combine to form a quasi-Dirac \cite{Valle:1983yw} or
pseudo-Dirac \cite{Wolfenstein:1981kw} neutrino, whose splitting
accounts for atmospheric oscillations, while the oscillations among
the two low-lying states explain the solar neutrino data, with the
overall scale accounting for the LSND result.

Although less motivated theoretically, it has been argued recently
that (3+1) schemes are not strictly ruled out
phenomenologically~\cite{barger00,carlo,peres}. However, using the
most recent atmospheric neutrino data, as well as short-baseline and
tritium data we show that such schemes are severely disfavored. In
order to accomplish this we extend the analysis of neutrino
oscillation data in the framework of (3+1) neutrino mass spectra
performed in Ref.~\cite{GS}. In addition to the full data from the
short-baseline (SBL) experiments Bugey~\cite{bugey}, CDHS~\cite{CDHS},
KARMEN~\cite{KARMEN2000} and the result of the long-baseline reactor
experiment CHOOZ~\cite{CHOOZ}, we also include the full and updated
data set of atmospheric neutrino experiments and the data from the
$\nu_\mu\to\nu_e$ oscillation search in NOMAD~\cite{nomad}. With this
information we derive a bound on the LSND amplitude $A_{\mu;e}$ within a
Bayesian statistical framework.
We find that the inclusion of the full atmospheric zenith-angle
distribution data considerably strengthens the bound on the LSND
amplitude $A_{\mu;e}$ for low $\Delta m^2$, whereas the NOMAD data
strengthens the bound for high $\Delta m^2$ values.
In contrast, the details of the solar data do not play a key role in
the analysis, so that there is no need to perform a full-fledged
global fit of solar data.  Likewise the most recent SNO
results~\cite{Ahmad:2001an} will hardly affect our results.
Finally, we also perform a different statistical analysis of the data
in order to also include information from the tritium $\beta$-decay
experiments~\cite{troitsk,mainz}. This sets additional strong bounds on
4--neutrino spectra of the type (3+1)B (see Fig.~\ref{fig:4spectra}).

In contrast to the (3+1) schemes, the intrepretation of the LSND
anomaly in terms of (2+2) spectra is in good agreement with SBL
data~\cite{GS,BGG,OY,BGGS,barger98}. It is a general prediction of
(2+2) spectra that the sterile neutrino must take part either in solar
or atmospheric neutrino oscillations (or both \cite{Hirsch:2000a}).
Atmospheric neutrino data prefer $\nu_\mu \to \nu_\tau$ oscillations over
oscillations into sterile neutrinos (see \eg\ 
Refs.~\cite{Fornengo:2000sr,SKconf}). Also fits to solar neutrino data
in terms of active neutrino oscillations are better than sterile
neutrino oscillation fits (see \eg\ 
Refs.~\cite{SNOfits,Gonzalez-Garcia:2000aj}), especially after the
first SNO results~\cite{Ahmad:2001an}.  However, a joint analysis of
both solar and atmospheric neutrino data in a 4--neutrino framework
shows that an acceptable fit can be obtained for the (2+2)
schemes~\cite{GGMPG}.
 
This paper is organized as follows. In Sec.~\ref{sec:notation} we fix our
notations. In Sec.~\ref{sec:atm} the fit of atmospheric neutrino data in the
framework of (3+1) mass spectra and its implications for parameters relevant
in short-baseline oscillation experiments are discussed. In
Sec.~\ref{sec:bound} we present the bound on the LSND amplitude $A_{\mu;e}$
whereas in Sec.~\ref{sec:tritium} implications of tritium $\beta$-decay
experiments are discussed. In Sec.~\ref{sec:conclusions} we draw our
conclusions.

\section{Notation}
\label{sec:notation}

The Standard Model can be extended with an arbitrary number of singlet
Majorana leptons, as they carry no gauge anomalies
\cite{Schechter:1980a}. The minimal case is to have simply one single
neutrino~\cite{Schechter:1980b} which we assume to remain light (due,
for example, to some symmetry) and therefore able to take part in the
oscillations\footnote{The number of such light sterile neutrinos may
  also be constrained by primordial nucleosynthesis, see \eg\ 
  Refs.~\cite{OY,BBN,BBN4}.}.  In any such 4--neutrino gauge scheme
the charged current weak interaction is characterized by the lepton
mixing matrix $K_{\alpha j}$. This is a rectangular $3\times 4$ matrix arising
from the unitary $4\times 4$ neutrino mixing matrix diagonalizing the
neutrino mass matrix and the corresponding unitary $3\times 3$ matrix
diagonalizing the left-handed charged leptons. This matrix $K_{\alpha j}$
contains in general six mixing angles and six CP
phases~\cite{Schechter:1980a}.

All neutrino oscillation probabilities in vacuo are determined by the
structure of the matrix $K_{\alpha j}$. For the case of solar and atmospheric
neutrinos matter effects in the solar and/or Earth interiors must also be
taken into account.

The probability of SBL $\pnu{\mu} \to \pnu{e}$ transitions relevant for the
accelerator experiments LSND, KARMEN and NOMAD is given by a very simple
two-neutrino-like formula~\cite{BGG}
\begin{equation} \label{eq:app}
    P_{\nu_\mu\to\nu_e} = P_{\bar\nu_\mu\to\bar\nu_e}
    = A_{\mu;e} \; \sin^2 \frac{\Delta m^2 L}{4E}
    \qquad \mathrm{with} \quad
    A_{\mu;e} = 4\, d_e d_\mu
\end{equation}
where $L$ is the distance between source and detector and $E$ is the neutrino
energy and the parameters $d_\alpha$ are defined as
\begin{equation}
    d_\alpha = |K_{\alpha 4}|^2 \quad (\alpha=e,\mu)\,.
\end{equation}
Note that for SBL oscillations we can safely neglect solar and
atmospheric splittings relative to the LSND gap. With our labeling of
the neutrino masses indicated in Fig.~\ref{fig:4spectra} the mass
separated by the LSND gap is denoted by $m_4$. This is the heaviest
mass in spectra of type (3+1)A and the lightest in (3+1)B spectra. As
a result $\Delta m^2_\mathrm{SBL} \equiv \Delta m^2 \approx |m^2_4-m^2_1|$ in all cases.
The LSND experiment gives an allowed region in the $(\Delta m^2, A_{\mu;e})$
plane.

The survival probabilities relevant in the SBL disappearance experiments Bugey
and CDHS are given by
\begin{equation} \label{eq:disapp}
    P_{\nu_\alpha\to\nu_\alpha} = 
    P_{\bar\nu_\alpha\to\bar\nu_\alpha} =
    1 - 4\, d_\alpha (1-d_\alpha) \sin^2 \frac{\Delta m^2 L}{4E} \,,
\end{equation}
where $\alpha = e$ refers to the Bugey and $\alpha = \mu$ to the CDHS experiment.
The result of the Bugey experiment requires $d_e$ to be very small or
very close to 1.  One can show that for the (3+1) spectra the survival
probability of solar neutrinos is bounded by $P_{\nu_e\to\nu_e}^\odot \geq
d_e^2$~\cite{BGG}, so that $d_e$ must be small, and we can include
this information from solar neutrinos in the analysis through the
approximation $d_e(1-d_e)\approx d_e$~\cite{GS}.

\section{Atmospheric data and short-baseline oscillations}
\label{sec:atm}

We now discuss the implications of atmospheric neutrino data on SBL
oscillation parameters. In Ref.~\cite{BGGS} it was shown that the
up-down asymmetry of atmospheric multi-GeV $\mu$-like events measured in
the Super-Kamiokande experiment~\cite{SK-atm-98} can be used to
constrain the parameter $d_\mu$ to values smaller than around 0.5. In
the following we will see that a detailed fit to the full atmospheric
neutrino data gives a much stronger constraint on $d_\mu$.

For the analysis of atmospheric neutrinos we use the latest
experimental data used in Ref.~\cite{GGMPG}: $e$-like and $\mu$-like
data samples of sub- and multi-GeV and up-going muon data including the
stopping and through-going muon fluxes from
Super-Kamiokande~\cite{SKconf} and the latest MACRO~\cite{macro}
up-going muon samples. For further details see
Refs.~\cite{Gonzalez-Garcia:2000aj,GGMPG,GMPV}.

The analysis of atmospheric neutrino data presented in
Ref.~\cite{GGMPG} was performed for the (2+2) spectra adopting the
approximations $\Delta m_\odot^2 = 0\,$ and $\Delta m_\mathrm{LSND}^2 \to \infty$.
Moreover, in that analysis the electron neutrino was considered as
completely decoupled from the atmospheric oscillations. This
approximation is well justified for (2+2) spectra, because in this
case the projection of $\nu_e$ over the atmospheric states is severely
restricted by the very strong Bugey bound. 
In contrast, in (3+1) spectra the contribution of electron neutrinos
to atmospheric oscillations is limited only by the somewhat weaker
CHOOZ bound.  However, in Ref.~\cite{GMPV} it was found that a $\nu_e$
contamination small enough not to spoil the results of the CHOOZ
experiment has only a very small effect on the quality of the fit of
atmospheric neutrino data. Therefore, even in the context of (3+1)
schemes it is reasonable to assume that electron neutrinos are
decoupled, so that atmospheric oscillations actually involve only
three neutrino flavours ($\nu_\mu$, $\nu_\tau$ and $\nu_s$). Under these
approximations, the (2+2) and (3+1) spectra become identical (except
for irrelevant signs), and therefore it is possible to use the
analysis given in Ref.~\cite{GGMPG} also in the context of (3+1)
schemes. In particular, the parameter $d_\mu$ used in the present work
corresponds to the parameter $s_{23}^2 = |U_{\mu 1}|^2 + |U_{\mu 2}|^2$ of
Ref.~\cite{GGMPG}.

\begin{figure}[t] \centering
    \includegraphics[height=8cm]{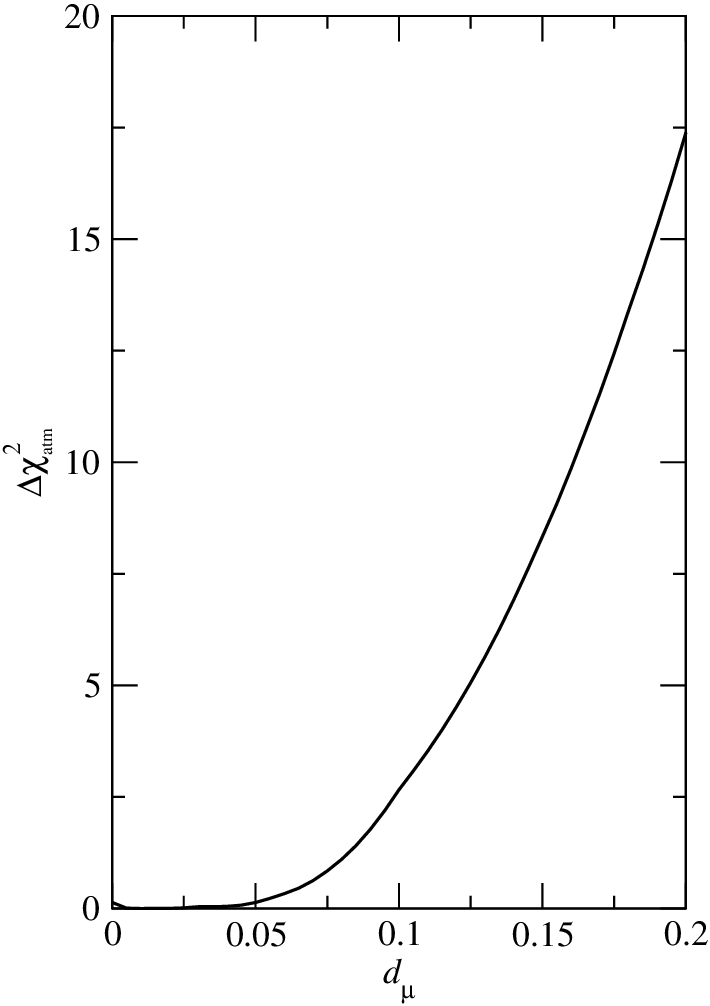} \hfil
    \includegraphics[height=8cm]{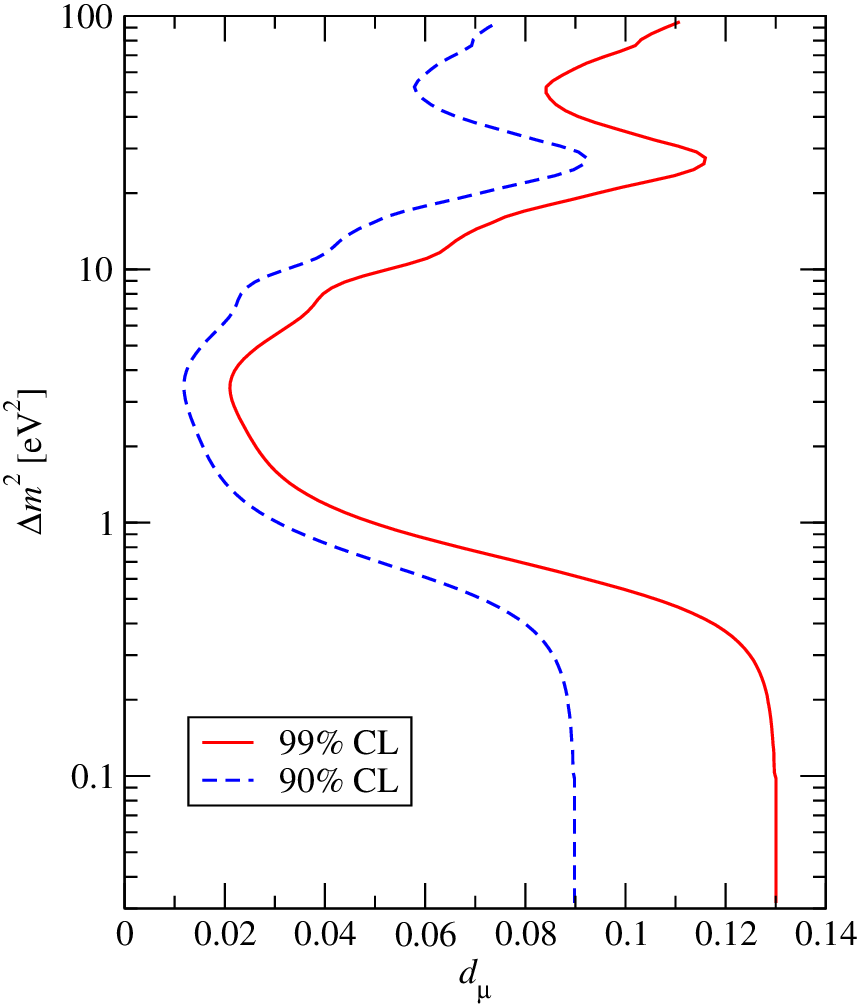} 
    \caption{\label{fig:atmData} %
      Left panel: $\Delta \chi^2_\mathrm{atm}$ as a function of $d_\mu$. Right
      panel: 90\% and 99\% CL upper bounds on $d_\mu$ by combining atmospheric
      neutrino data and the CDHS experiment. }
\end{figure}
 
In the left panel of Fig.~\ref{fig:atmData} we show $\Delta\chi^2_\mathrm{atm}
= \chi^2_\mathrm{atm} - (\chi^2_\mathrm{atm})_\mathrm{min}$ from the fit to
atmospheric neutrino data as a function of the parameter $d_\mu$.  For
each value of $d_\mu$ the $\chi^2$ is minimized with respect to all other
undisplayed parameters necessary to fit the atmospheric neutrino data.
In the right panel of Fig.~\ref{fig:atmData} we show the 90\% and 99\%
CL bounds on $d_\mu$ obtained from a combination of all the atmospheric
neutrino data with the $\nu_\mu$ disappearance experiment CDHS in a
Bayesian framework\footnote{An analysis using a $\chi^2$-cut method gives
  very similar results.}.

In the lower part of this plot the constraint on $d_\mu$ comes from
atmospheric neutrino data alone, as the CDHS bound disappears for $\Delta
m^2 \lesssim 0.3$ eV$^2$. Hence, atmospheric neutrino data leads to the bound
\begin{equation} \label{eq:dmuBound}
    d_\mu \leq 0.090\,(0.13)\quad\mathrm{at}\quad 90\% \,(99\%)\,
    \mathrm{CL}\,.
\end{equation}
Let us note that Fig.~\ref{fig:atmData} and the bound given in
Eq.~\eqref{eq:dmuBound} are valid also for the (2+2) spectra if the parameter
$d_\mu$ is identified with $|K_{\mu 1}|^2 + |K_{\mu 2}|^2$ and neutrino masses
are labeled according to Fig.~\ref{fig:4spectra} (right).  

\section{An upper bound on $A_{\mu;e}$}
\label{sec:bound}

In this section we discuss the upper bound on the LSND amplitude $A_{\mu;e}$
obtained by combining the data of the SBL experiments Bugey, CDHS, KARMEN and
NOMAD, with those of the atmospheric neutrino experiments and the CHOOZ
experiment. Here we focus mainly on the results of the extendend analysis,
technical details can be found in Ref.~\cite{GS}.

To combine all the oscillation data we use the likelihood function
\begin{equation} \begin{split} \label{eq:LH}
    \mathcal{L}_\mathrm{osc}(d_e, d_\mu, \Delta m^2) 
    & = \mathcal{L}_\mathrm{Bugey}(d_e, \Delta m^2) \, 
    \mathcal{L}_\mathrm{CDHS}(d_\mu, \Delta m^2) \\
    & \times \mathcal{L}_\mathrm{KARMEN}(d_ed_\mu, \Delta m^2) \,
    \mathcal{L}_\mathrm{NOMAD}(d_e d_\mu, \Delta m^2) \\
    & \times \mathcal{L}_\mathrm{atm}(d_\mu) \,
    \mathcal{L}_\mathrm{CHOOZ}(d_e) \,.
\end{split} \end{equation}
The likelihood functions $\mathcal{L}_\mathrm{Bugey}$,
$\mathcal{L}_\mathrm{CDHS}$ and $\mathcal{L}_\mathrm{KARMEN}$ are described in
Ref.~\cite{GS} and $\mathcal{L}_\mathrm{atm} \propto \exp\left(
-\frac{1}{2}\chi^2_\mathrm{atm} \right)$~\cite{PDG00}. To calculate
$\mathcal{L}_\mathrm{NOMAD}$ we perform a re-analysis of the $\nu_\mu\to\nu_e$
oscillation search at NOMAD by using the data given in Ref.~\cite{nomad}. The
result of the CHOOZ experiment is included via $\mathcal{L}_\mathrm{CHOOZ}$,
which is obtained with the maximum likelihood method described also in
Ref.~\cite{GS}.

\begin{figure}[t] \centering
    \includegraphics[width=0.85\linewidth]{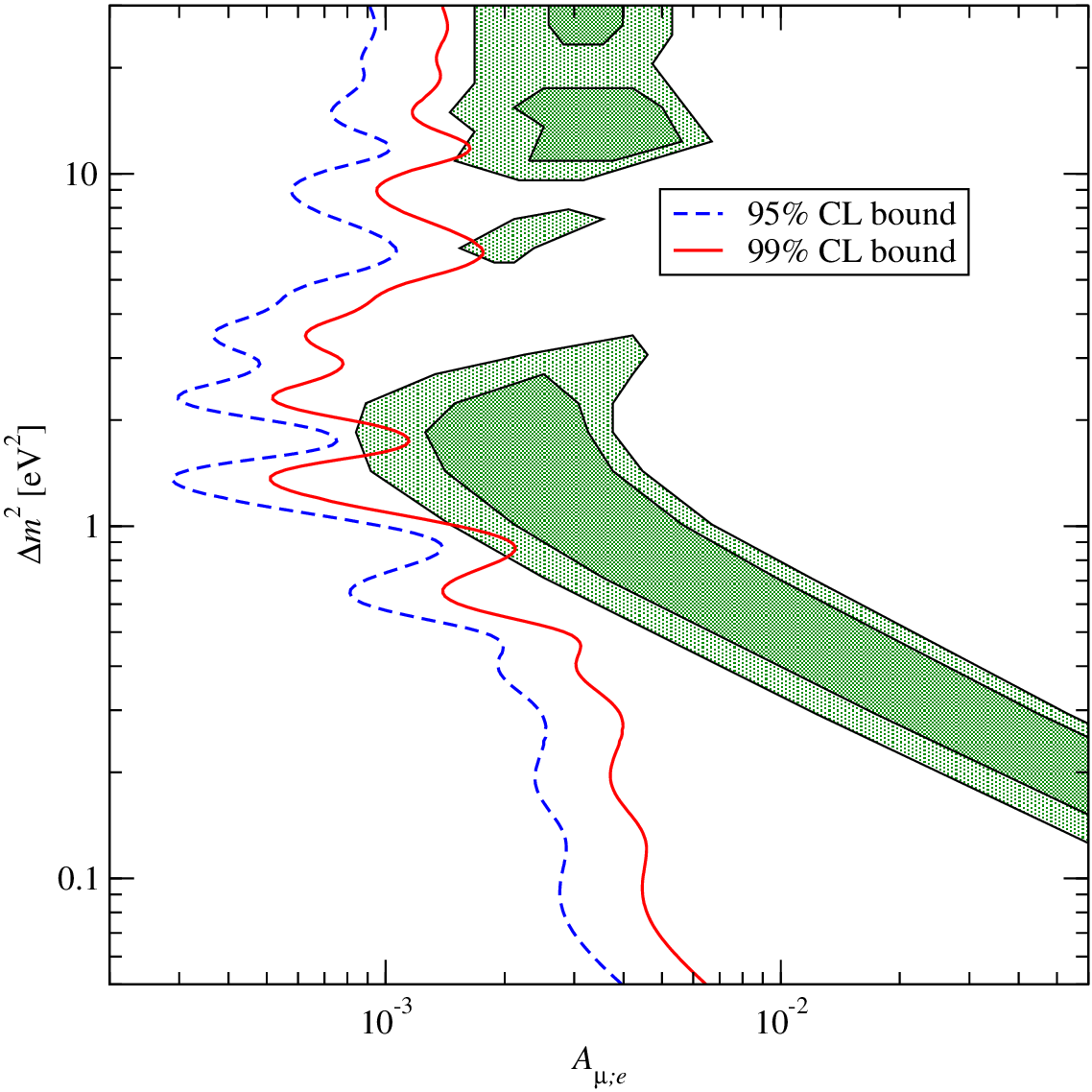}
    \caption{\label{fig:Abound}%
      Upper bounds on the LSND amplitude $A_{\mu;e}$ at 95\% and 99\% CL. The
      shaded regions are the regions allowed by LSND at 90\% and 99\%
      CL~\cite{LSND2000}.}
\end{figure}

For a {\it fixed} value of $\Delta m^2$ the likelihood function
Eq.~\eqref{eq:LH} is transformed into a probability distribution
$p(d_e,d_\mu)$ by applying Bayes' theorem (see for example
Refs.~\cite{PDG00,cowan}) and assuming a flat prior in the physically allowed
region $d_e,d_\mu \geq 0$ and $d_e + d_\mu \leq 1$.  Choosing a CL $\beta$, we
find the corresponding upper bound $A^0_\beta$ on $A_{\mu;e}$ 
by the prescription
\begin{equation}\label{eq:Abound}
    \int\limits_{4d_ed_\mu \leq A^0_\beta} \hskip-10pt
    \mathrm{d}d_e\, \mathrm{d}d_\mu\, p(d_e,d_\mu) = \beta \,.
\end{equation}
The bounds at 95\% and 99\% CL are shown in Fig.~\ref{fig:Abound} together
with the regions allowed by LSND at 90\% and 99\% CL.  We find that there is
no overlap of the region allowed by our bound at 95\% CL with the region
allowed by LSND at 99\% CL~\cite{GS}. If we take our bound at 99\% CL there
are marginal overlaps with the 99\% CL LSND allowed region at $\Delta m^2 \sim
0.9$ and 2 eV$^2$, and a very marginal overlap region still exists around 6
eV$^2$.  The overlap region found in Ref.~\cite{GS} between 0.25 and 0.4
eV$^2$ is now excluded by our bound at 99\% CL because of the inclusion of the
full atmospheric neutrino data set\footnote{We note that for $\Delta m^2
\gtrsim 10$ eV$^2$ there are additional constraints on the amplitude
$A_{\mu;e}$ from the experiments BNL E776~\cite{BNL} and
CCFR~\cite{CCFR}, which are not included in our analysis. Therefore, the (in
any case very marginal) overlap region at $\Delta m^2 \sim 10$ eV$^2$ in
Fig.~\ref{fig:Abound} is irrelevant.}.

For the alternative case of symmetric (2+2) spectra the bound on the
LSND amplitude $A_{\mu;e}$ is dominated by the Bugey bound on $d_e$ and
by the KARMEN bound on $A_{\mu;e}$. Hence, an improvement of the bound
on $d_\mu$ by the inclusion of the full atmospheric neutrino data does
not lead to a stronger bound on $A_{\mu;e}$ in this case so that the
results of Ref.~\cite{GS} apply.  The conclusion is that (2+2) spectra
are in good agreement with SBL data (see Fig.~2 of Ref.~\cite{GS}). As
shown in \cite{GGMPG} they are also in agreement with the totality of
solar and atmospheric neutrino data.

\section{Implications of tritium $\beta$-decay}
\label{sec:tritium}

Experiments studying the electron spectrum $\mathrm{d}N/\mathrm{d}E_e$
from tritium $\beta$-decay can obtain information on the quantity $m_\beta^2$
determined by the relation
\begin{equation} \label{eq:mbeta2}
    \frac{\mathrm{d} N}{\mathrm{d} E_e} \propto
    \sqrt{(E_e-E_0)^2 - m_\beta^2} \,,
\end{equation}
where $E_e$ is the energy of the electron and $E_0$ is the total decay energy.
The latest result obtained by the Troitsk experiment is~\cite{troitsk}
\begin{equation} \label{eq:v1}
    m_\beta^2 = -1.0 \pm 3.0 \pm 2.1 \,\mathrm{eV}^2\,
    ,\qquad m_\beta < 2.5 \,\mathrm{eV}\,.
\end{equation}
The Mainz collaboration recently presented two values, obtained from different
analyses of their data~\cite{mainz}:
\begin{align}
    \label{eq:v2} m_\beta^2 &= +0.6 \pm 2.8 \pm 2.1 \, \mathrm{eV}^2 \,
    ,\qquad m_\beta < 2.8 \,\mathrm{eV}\,, \\
    \label{eq:v3} m_\beta^2 &= -1.6 \pm 2.5 \pm 2.1 \, \mathrm{eV}^2 \,
    ,\qquad m_\beta < 2.2 \,\mathrm{eV}\,.
\end{align}
In Eqs.~\eqref{eq:v1}--\eqref{eq:v3} the upper bounds are at 95\% CL.

Let us now consider the implications of these measurements for the (3+1)A
and (3+1)B neutrino mass schemes (see Fig.~\ref{fig:4spectra}).
In the presence of neutrino mixing relation (\ref{eq:mbeta2})
has to be modified to (see \eg~\cite{farzan})
\begin{equation}\label{eq:mbeta2m}
    \frac{\mathrm{d} N}{\mathrm{d} E_e} \propto
    \sum_{i=1}^4 |K_{ei}|^2 \,\sqrt{(E_e-E_0)^2 - m_i^2} \,
    \Theta(E_0-E_e-m_i) \,.
\end{equation}
In view of the 
very strong constraint on $d_e$ from Bugey we  can safely neglect
the contribution from $|K_{e4}|^2$ to the sum in Eq.~(\ref{eq:mbeta2m}).
Further we take into account that mass splittings implied by solar and
atmospheric neutrino data cannot be resolved in tritium decay experiments
\cite{farzan}. Hence, in spectra (3+1)A the value $m_\beta$ is given by the
lowest neutrino mass and is independent of $\Delta m^2$ to a good 
approximation. Therefore, we do not include any information from 
tritium $\beta$-decay in this case. However, for spectra of the type 
(3+1)B one obtains \cite{BPP} 
$m_\beta^2 \approx m_4^2 + \Delta m^2$.\footnote{Note that in our 
notation in scheme (3+1)B the lightest neutrino mass is $m_4$.}
%
%
We include this result in our statistical analysis by using the likelihood
function
\begin{equation} \label{eq:lhbeta}
    \mathcal{L}_\beta(\Delta m^2) \propto 
    \begin{cases}
        \mathrm{const} & \text{for (3+1)A} \\ 
        \exp \left[ -\frac{1}{2} \sum\limits_i \left( 
        \frac{(m_\beta^2)_i - {\Delta m^2} - {m^2_4}}
             {\sigma_i} \right)^2 \right]
        & \text{for (3+1)B}
    \end{cases}
\end{equation}
where the sum is over the three experimental values of $m_\beta^2$ given in 
Eqs.~\eqref{eq:v1}--\eqref{eq:v3} and $\sigma_i$ is the corresponding error
(statistical and systematic errors added in quadrature). 
For fixed values of $m_4$ we perform now
an analysis with the two parameters $A_{\mu;e}$ and $\Delta m^2$, in contrast
to Sec.~\ref{sec:bound}, where the analysis is performed only with the
parameter $A_{\mu;e}$ for each value of $\Delta m^2$. 

As a first step the total likelihood function obtained from Eqs.~\eqref{eq:LH}
and~\eqref{eq:lhbeta}
\begin{equation}
    \mathcal{L}_\mathrm{tot}(d_e,d_\mu,\Delta m^2) =
    \mathcal{L}_\beta(\Delta m^2) \, 
    \mathcal{L}_\mathrm{osc}(d_e,d_\mu,\Delta m^2)
\end{equation}
is transformed into a probability distribution $p(d_e,d_\mu,\log\Delta m^2)$
by using Bayes' theorem. We assume a flat prior distribution for $d_e$ and
$d_\mu$ in the physical region and a flat prior distribution for $\log\Delta
m^2$. This ensures that we introduce no bias concerning the order of magnitude
of $\Delta m^2$, a priori all scales are equally probable.  Then we perform a
transformation of the variables $d_e$ and $d_\mu$ to\footnote{Note that the
Jacobi determinant of this transformation is 1, hence
$\mathrm{d}d_e\mathrm{d}d_\mu = \mathrm{d}A_{\mu;e}\mathrm{d}t$.}
\begin{equation}
    A_{\mu;e} = 4\,d_e d_\mu \,,\quad t=\frac{1}{8}\ln\frac{d_\mu}{d_e}
\end{equation}
and integrate over the variable $t$ to obtain the probability distribution for
the variables we are interested in: $p(A_{\mu;e},\log\Delta m^2)$. We
calculate an allowed region at the 100$\beta$\% CL by demanding
\begin{equation}
    \int \mathrm{d}A_{\mu;e}\mathrm{d} (\log\Delta m^2) \, 
    p(A_{\mu;e},\log\Delta m^2) = \beta\,.
\end{equation}
The boundary in the $(A_{\mu;e}, \Delta m^2)$ plane is determined such that
the value of $p(A_{\mu;e},\log\Delta m^2)$ along this line is constant.

\begin{figure}[t] \centering
    \includegraphics[width=0.85\linewidth]{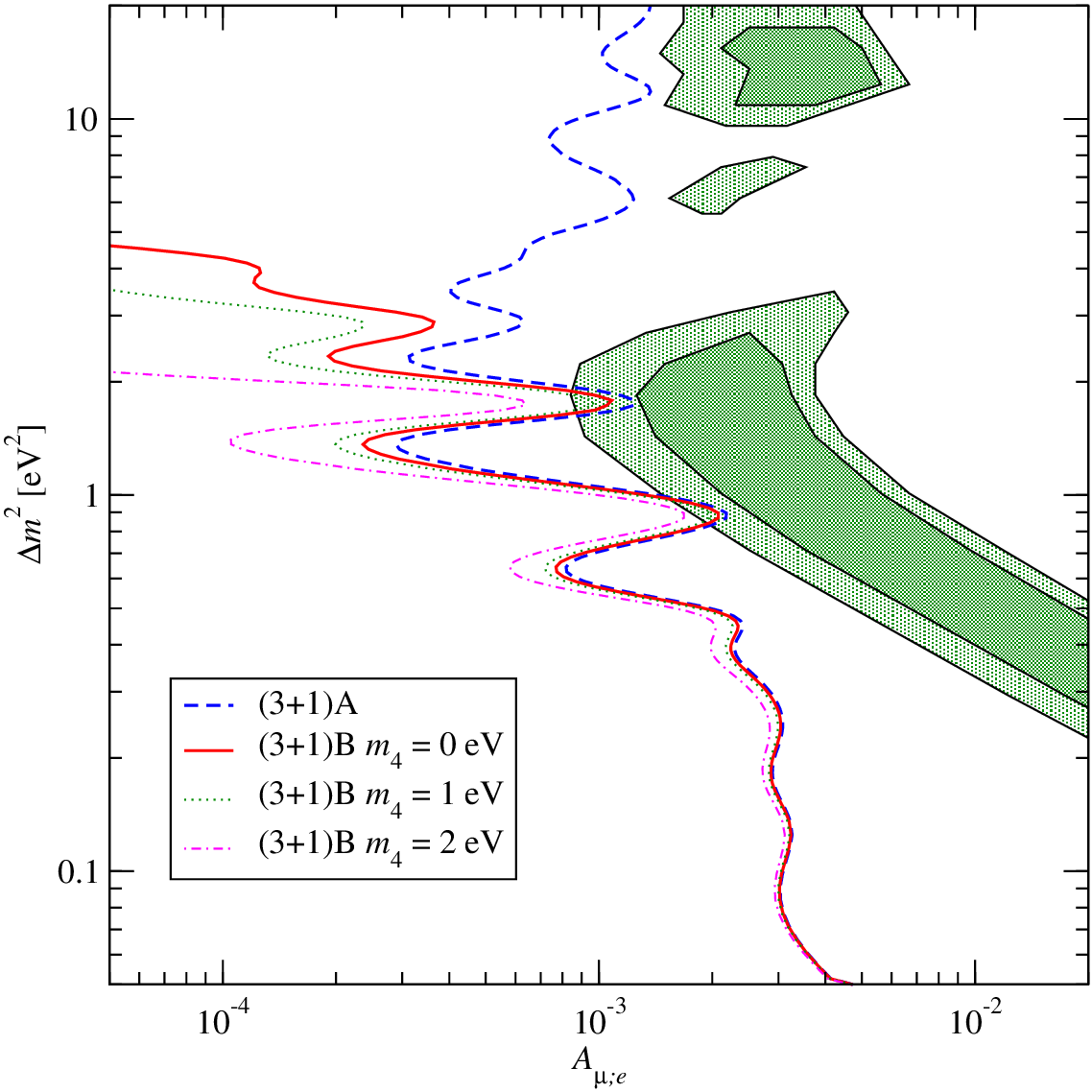}
    \caption{\label{fig:tritBound}%
      Allowed regions at 99\% CL in the $(A_{\mu;e}, \Delta m^2)$ plane for
      spectra of the types (3+1)A and (3+1)B including tritium $\beta$-decay.
      In case of (3+1)B we show the regions for values
      of the lightest neutrino mass $m_4 = 0$, 1 and 2 eV.
      The shaded regions are the regions allowed by LSND at 90\% and 99\%
      CL~\cite{LSND2000}.}
\end{figure}

In Fig.~\ref{fig:tritBound} we show the allowed regions at 99\% CL for
spectra of the types (3+1). In the case of (3+1)B we show the regions 
for $m_4 = 0$, 1 and 2 eV. As we do not know the true value of
$m_4$ the curve corresponding to vanishing $m_4$ is the most
conservative one. In this case
tritium $\beta$-decay rules out values of 
$\Delta m^2 \gtrsim 5$ eV$^2$. In both
cases (3+1)A and (3+1)B ($m_4=0$)
only two marginal overlaps with the 99\%CL LSND
allowed region survive at $\Delta m^2 \sim 0.9$ and $2$ eV$^2$.
For (3+1)B with $m_4=2$ eV the overlap with LSND disappears. 

The following remarks are in order:
\begin{enumerate}
  \item To normalize the probability distribution $p(A_{\mu;e}, \log\Delta
    m^2)$ one has to choose a lower integration bound for $\log\Delta m^2$ as
    $p(A_{\mu;e},\log\Delta m^2)$ does not vanish for $\log\Delta m^2 \to
    -\infty$. The allowed regions in Fig.~\ref{fig:tritBound} depend somewhat
    on this lower bound. However, if one chooses the lower bound sufficiently
    small ($\log\Delta m^2 \lesssim -3$) the allowed regions become
    independent of it. In the case of the spectra (3+1)A the allowed region
    depends also on the upper integration bound for $\log\Delta m^2$. However,
    again the dependence disappears, if this bound is chosen large enough
    ($\log\Delta m^2 \gtrsim 2$).
    
  \item Note that the statistical meaning of the bounds in
    Figs.~\ref{fig:Abound} and~\ref{fig:tritBound} is different. The
    method applied in Sec.~\ref{sec:bound} to produce
    Fig.~\ref{fig:Abound} allows to place an upper bound on $A_{\mu;e}$
    for a {\it given} value of $\Delta m^2$ at a certain CL, whereas the
    meaning of the 99\% CL bounds shown in Fig.~\ref{fig:tritBound} is
    the following: the true values of $A_{\mu;e}$ {\it and} of $\Delta m^2$
    are expected to lie at the left of the curves shown in the figure
    with probability 0.99. This explains the small differences between
    the curve for the (3+1)A spectra in Fig.~\ref{fig:tritBound} and
    the 99\% CL curve in Fig.~\ref{fig:Abound}.  However, the general
    agreement of both methods renders confidence to our analysis.

  \item In our analysis we take into account only the shape of the
    likelihood as a function of $\Delta m^2$ for fixed values of
    $m_4$; for each given value of $m_4$ the likelihood function is
    normalized to 1. However, because of the relation
    $m_\beta \approx \sqrt{m_4^2 + \Delta m^2}$ in scheme (3+1)B
    there are additional strong constraints on the allowed values
    of $\Delta m^2$ and $m_4$ from the upper bounds on $m_\beta$
    given in Eqs.~\eqref{eq:v1}--\eqref{eq:v3}.
%
\end{enumerate}

Finally, we note that the non-observation of neutrinoless double 
$\beta$-decay may also place important restrictions on (3+1)B
spectra~\cite{BPP}, where the electron neutrino has a substantial
component along the heaviest neutrinos. However, this will be very
model dependent as the resulting bounds are subject to possible
destructive interference due to cancellations among different
neutrinos \cite{Valle:1983yw,Wolfenstein:1981rk}.

\section{Conclusions}
\label{sec:conclusions}

We have extended the analysis of neutrino oscillation data in the
framework of (3+1) neutrino mass spectra performed in Ref.~\cite{GS}.
In addition to the full data from the short-baseline experiments
Bugey~\cite{bugey}, CDHS~\cite{CDHS}, KARMEN~\cite{KARMEN2000} and the
result of the long-baseline reactor experiment CHOOZ~\cite{CHOOZ}, we
have included also the full and updated data set of atmospheric
neutrino experiments, the data from the $\nu_\mu\to\nu_e$ oscillation search
in NOMAD~\cite{nomad} and tritium $\beta$-decay data~\cite{troitsk,mainz}.
We have shown that the interpretation of the LSND anomaly within such
schemes is severely disfavored by the combined data, in contrast to
the case of the theoretically preferred (2+2) schemes. Since the
details of the solar data do not play an important role in our (3+1)
analysis, the most recent SNO results~\cite{Ahmad:2001an} will not
affect the conclusions derived in our paper, nor contribute to enhance
the bounds we have derived. Our present results put an additional
challenge on some recent attempts to revive (3+1)
schemes~\cite{barger00,carlo,peres}.

\section*{Acknowledgements}

We would like to thank W.~Grimus and C.~Pe{\~n}a-Garay for very useful
discussions. This work was supported by Spanish DGICYT under grant
PB98-0693, by the European Commission RTN network HPRN-CT-2000-00148
and by the European Science Foundation network grant N.~86. T.S.\ was
supported by a fellowship of the European Commission Research Training
Site contract HPMT-2000-00124 of the host group. M.M.\ was supported
by contract HPMF-CT-2000-01008.

\end{document}